\documentclass[twocolumn,preprintnumbers,amsmath,amssymb]{revtex4}
\usepackage{amsmath}
\usepackage{graphicx}
\newcommand\etal{\mbox{\textit{et al.}}}
\begin{document}
%\doi{10.1080/14685240xxxxxxxxxxxxx}
%\issn{1468-5248}
%\issnp{} \jvol{00} \jnum{00} \jyear{2008} \jmonth{September}

%\markboth{W.J.T. Bos and J.-P. Bertoglio}{Large-scale bottleneck effect in two-dimensional turbulence}

\title{\large Large-scale bottleneck effect in two-dimensional turbulence}

\author{Wouter J.T. Bos\thanks{wouter.bos@ec-lyon.fr},
and Jean-Pierre Bertoglio\\
\vspace{0.2cm}
Universit\'e de Lyon - LMFA - CNRS UMR 5509 \\
 Ecole Centrale de Lyon - 69134 Ecully, France \\
 INSA de Lyon, Universit\'e Lyon 1 
}
%\received{September 2008}

\begin{abstract}
{\bf Abstract.} The bottleneck phenomenon in three-dimensional turbulence is generally
associated with the dissipation range of the energy spectrum. In the present work, it is shown by using a two-point closure theory,  that in two-dimensional turbulence it is possible to observe a bottleneck at the large scales, due to the effect of friction on the inverse energy cascade.
This large-scale bottleneck is directly related to the
process of energy condensation, the pile up of energy at wavenumbers
corresponding to the domain size. The link between the use of friction
and the creation of space-filling structures is discussed and  
it is concluded that the careless use of hypo-friction might reduce the inertial range of the energy spectrum.
\bigskip

\end{abstract}

\maketitle

\section{Introduction}

In high Reynolds number three-dimensional turbulence at large
wavenumbers $k$, the compensated energy spectrum $k^{5/3}E(k)$ often
shows a little positive bump, before it drops down rapidly in the
dissipation range. This pre-dissipative bump region was first
observed in closure computations \cite{Andre1977}, but no physical
explanation was given. The bump was also observed in experimental
work \cite{Mestayer1982,Saddoughi} and direct numerical simulations
\cite{Kerr1990}. A physical explanation of the effect was first
given in the work by Herring \etal \cite{Herring}. They explain
that, because the wavenumbers above the Kolmogorov scale are damped
by viscosity, some of the nonlocal interactions are damped, which
causes the transfer of the energy flux to be less effective, leading
to a pile-up of energy. A more detailed investigation of this effect
was performed by Falkovich \cite{Falkovich}, who predicted its
wavenumber-dependence, and who named it the \emph{bottleneck}. In
that work it is also inferred that the bottleneck should become more
pronounced when hyperviscosity is used. This was illustrated in
three-dimensional turbulence\cite{Borue1995} and two-dimensional
turbulence with a direct energy cascade \cite{Biskamp1998} such as
magnetohydrodynamic turbulence. 

In the present work, we show that in two-dimensional turbulence it is
possible to observe a bottleneck at the \emph{large} scales, due to
the effect of friction on the inverse energy cascade. 
The possibility
of large-scale bottlenecks was advanced by Lohse and Mueller-Groeling
\cite{Lohse1995}. This range is in three-dimensional turbulence
generally affected by a forcing which supplies energy to the flow,
complicating the observation of a possible large-scale bottleneck.

An alternative interpretation of the bottleneck in three-dimensional
turbulence was
recently proposed by Frisch and coworkers \cite{Frisch2008}. They
relate the bottleneck to statistical equilibrium, and view it as an
incomplete thermalization. Indeed, it was stated that if the
Laplacian $\nabla^2$ in the viscous term of the Navier-Stokes
equations is replaced by a hyperviscous term $\nabla^{2\alpha}$,
with the \emph{dissipativity} $\alpha$ tending to infinity, a
Galerkin truncated system is obtained. In Galerkin-truncated
3D inviscid turbulence, energy piles up at the highest wavenumbers,
leading to wavemodes in equipartition at the tail of the spectrum.
The simultaneous observation of a $k^{-5/3}$ Kolmogorov inertial
range energy spectrum and this equipartition range, proportional to
$k^2$, was recently observed in direct numerical
simulations\cite{Cichowlas} (DNS)  and closure calculations
\cite{Bos2006-2}. In between these two ranges a pseudo-dissipation
range was observed, caused by nonlocal interactions between the
thermal bath and the end of the inertial range.

Since we are interested in the large scales in the present work, we
discuss the influence of infrared wavenumber truncation, due to the
finite domain-size. For three-dimensional turbulence, this truncation influences decaying turbulence and increases the energy decay rate, if the size of the integral lengthscale becomes comparable to the domain-size \cite{Touil2}. Steady turbulence, forced at a wavenumber $k_i>k_0$, in which $k_0$ is the smallest wavenumber of the domain, called the \emph{fundamental}, will establish an infrared region proportional to $k^2$  between $k_0$ and $k_i$, independent of the exact choice of $k_0$ \cite{Lesieur}.

In two space-dimensions, the effect of infrared truncation is very
different. The absence of vortex stretching induces a
double cascade with energy cascading to the large scales and
enstrophy going to the small scales. In this case, the large scales
are not directly influenced by forcing. The energy spectrum is not
necessarily decreasing for $k$ tending to zero, if no energy-sink is present at low $k$. In an infinite domain, the energy will continue to cascade to smaller and smaller wavenumbers. In the case of a finite domain-size, the energy, cascading backwards, piles up at
the fundamental. This  condensation was predicted by Kraichnan \cite{Kraichnan67} and observed numerically \cite{Hossain1983}. 
The condensation process was studied in detail by Smith and Yakhot \cite{Smith1993,Smith1994}, who showed that the statistics of the velocity field remain close to Gaussian until the condensate starts to form. At later times the statistics become non-Gaussian. We note that the condensate state is not the only possible state of finite-size two-dimensional turbulence in the absence of an infrared energy sink. Indeed Tran and Bowman \cite{Tran2003,Tran2004} showed that if the inverse energy cascade is weak, \emph{i.e.}, if only a small part of the injected energy cascades to small wavenumbers, the rest being dissipated around the forcing scale, no condensation takes place, but a gradual steepening of the spectrum occurs, when the cascade reaches the smallest wavenumbers, correponding to the box-size. This steepening results in a $k^{-3}$ range at late times. This scenario was also observed by Chertkov \emph{et al.} \cite {Chertkov2007}. In the present work we consider the case in which the energy cascade is strong and in which condensation takes place at the fundamental, when no energy sink is present.

\section{The link between hypofriction and energy condensation}

To avoid condensation in two-dimensional turbulence, large-scale friction is often used. Friction acts as an energy sink. The
influence of friction, which in principle acts throughout the
spectrum, can be concentrated at the smallest wavenumbers by using
hypofriction, analogous to hyperviscosity at the highest
wavenumbers. The Navier-Stokes equations in this case become:
\begin{eqnarray}\label{eqNS}
\frac{\partial \bm u}{\partial t}+(\bm u\cdot\nabla)\bm u=-\nabla p+\nu \nabla^2 \bm u+\bm f&-\mu k_{f}^{2\alpha}(-\nabla^{2})^{-\alpha}\bm u;\nonumber\\
\nabla\cdot \bm u=0, &
\end{eqnarray}
in which $\bm u$ is the velocity vector, $p$ the pressure, $\nu$ the
viscosity, $\bm f$ an isotropic forcing term localized in spectral space around
wavenumber $k_i$, $\mu$ the friction coefficient, $k_{f}$ the
friction wavenumber, and we shall call $\alpha$ the
\emph{frictionality}, in analogy with the dissipativity in
hyperviscous flows\cite{Frisch2008}.

In the presence of hypofriction, an energy sink is present at low
$k$ (e.g. \cite{Borue1994,Huang2001}). In this case a steep fall-off
of the energy spectrum at low $k$ can be created for large $\alpha$. If the solution of the
hypofrictional case converges to a Galerkin truncated case if
$\alpha\rightarrow \infty$, in analogy with the work of Frisch \etal
\cite{Frisch2008}, a pile-up of energy can be expected around $k_f$. 
In the case of linear friction ($\alpha=0$), no noticable bottleneck is observed (e.g. \cite{Paret1998}). The
spectral slope $d~ln(E(k))/d~ln(k)$, is a continuously decreasing
function of $k$ for $k_0 < k\ll k_i$ with $k_0$ the fundamental and
$k_i$ the wavenumber where energy is introduced by the forcing.

In the present study, using a closure model, we will systematically vary the order of the frictionality from $\alpha=0$ to $\alpha=128$. We will investigate the appearance of bottlenecks, its influence on the inertial range scaling and the convergence of this bottleneck to an infrared  wavenumber truncation.

Some evidence for large-scale bottlenecks can be found in
literature \cite{Borue1994,Huang2001}. In these studies hypofriction
was used with values $8$ and $5$ respectively. The compensated
energy spectra do show a pile up of energy at the small wavenumbers. A
link with the creation of coherent structures is made but no
systematic study of the influence of $\alpha$ on the large-scale
bottleneck was reported in these studies.

Our interpretation, in the light of the present study, is that the bottlenecks in these works are in
fact incomplete condensates, in analogy with the interpretation of Frisch
\etal \cite{Frisch2008}, that the bottleneck in 3D turbulence is an incomplete thermalization. The coherent structures that are observed are
closely related to the structures, or condensates, that would be observed in a truncated solution.

Even though numerical resources increase rapidly and two-dimensional
turbulence is computationally accessible at high Reynolds numbers, a
systematic DNS study of the influence of $\alpha$ would be very
expensive. Furthermore the small wavenumber range of an energy
spectrum converges more slowly to a statistically steady state than
the small scales. It was shown that for this kind of studies,
two-point closure is a convenient tool\cite{Bos2006-2,Frisch2008}.
For the case of two-dimensional turbulence, it was pointed out
\cite{Herring1985} that in the absence of coherent vortices, closure
approaches compare well to direct numerical simulation. The inverse
cascade is known to be free from large coherent structures and its
statistics are close to Gaussian \cite{Boffetta2000}. In direct
numerical simulations, if friction is
absent, the pile up of energy at the fundamental will eventually give
rise to the creation of two counter-rotating vortices of the size of
the domain.

In the long-time limit, the statistics become highly non-Gaussian
\cite{Smith1993} so that the use of the model, which are based on an expansion
about Gaussianity, will probably not be justified \cite{Kraichnan1989,Chen1989,Bertoglio1987}.
%}
The time before this happens is
supposed to be reasonably well predicted by closure. 
It is this time-interval, in which pile-up of
kinetic energy at small wavenumbers takes place, and which precedes the
formation of coherent structures, on which we concentrate in the
present work.

\section{Observation of large-scale bottlenecks}

We use the Eddy-Damped Quasi Normal Markovian (EDQNM) closure. It
was used to study two-dimensional turbulence by Leith \cite{Leith2}
and Pouquet \etal \cite{Pouquet}. In this closure, the evolution
equation for the energy spectrum is solved\begin{equation}
\left[\partial_t +\nu k^2+\mu
(k_f/k)^{2\alpha}\right]E(k,t)=T_{NL}(k,t)+F(k).
\end{equation}
The nonlinear transfer $T_{NL}$ is
\begin{eqnarray}\label{TNL}
T_{NL}(k,t)=\frac{4}{\pi}\iint_{\Delta}\Theta_{kpq}~\frac{xy-z+2z^3}{\sqrt{1-x^2}}\left[k^2pE(p,t)E(q,t) \right.\nonumber\\
\left. -kp^2E(q,t)E(k,t)\right]\frac{dpdq}{pq}.~~~~~~~~~
\end{eqnarray}
The forcing is a  
time-independent
energy input of rate $c$, localized at $k=k_i$, $F(k)=c~\delta(k-k_i)$.
In equation (\ref{TNL}), $\Delta$
is a band in $p,q$-space so that the three wave-vectors ${\bm{k},
\bm{p}, \bm{q}}$ form a triangle. $x,y,z$ are the cosines of the
angles opposite to $k,p,q$ in this triangle. The characteristic time
$\Theta_{kpq}$ is defined as:
\begin{equation}
\Theta_{kpq}=\frac{1-exp(-(\eta_k+\eta_p+\eta_q)\times
t)}{\eta_k+\eta_p+\eta_q},
\end{equation} 
in which $\eta$ is the eddy
damping. In the present study we use the following expression for
the damping coefficient
\begin{equation}\label{ED}
\eta_k=\lambda\sqrt{\int_0^k s^2E(s,t)dS}+\nu k^2+\mu (k_f/k)^{2\alpha}.
\end{equation}
In expression (\ref{ED}), three contributions to the eddy damping
are present. The first two are the standard contributions introduced
in the EDQNM closure. For $\lambda$ we use the value $0.4$
\cite{Lesieur1985}. The third contribution is an ad-hoc modification of the damping to take
into account the presence of friction at large scale. A rigorous
derivation of the modification induced by the friction as done for
the rotating case \cite{Bellet2006} or using an improved
eddy-damping \cite{Bos2006} is not attempted in the present work.
The numerical set-up is similar to the one in \cite{Bos2006-2}.  
The wavenumbers are spaced as $k_p=k_0 r^{p-1}$ with $k_0$ the fundamental and $r>1$ a constant which determines the number of wavenumbers per decade.
The spatial resolution for the computations reported here is $40$
wavenumbers per decade which corresponds to a resolution of
$F\approx 12$ wavenumbers per octave. Simulations are reported for
$\alpha=0,1,2,4,8,32,128$ as well as a friction-free simulation
truncated at $k_0=k_f$.

\begin{figure*}
\setlength{\unitlength}{1\textwidth}
\includegraphics[width=1\unitlength]{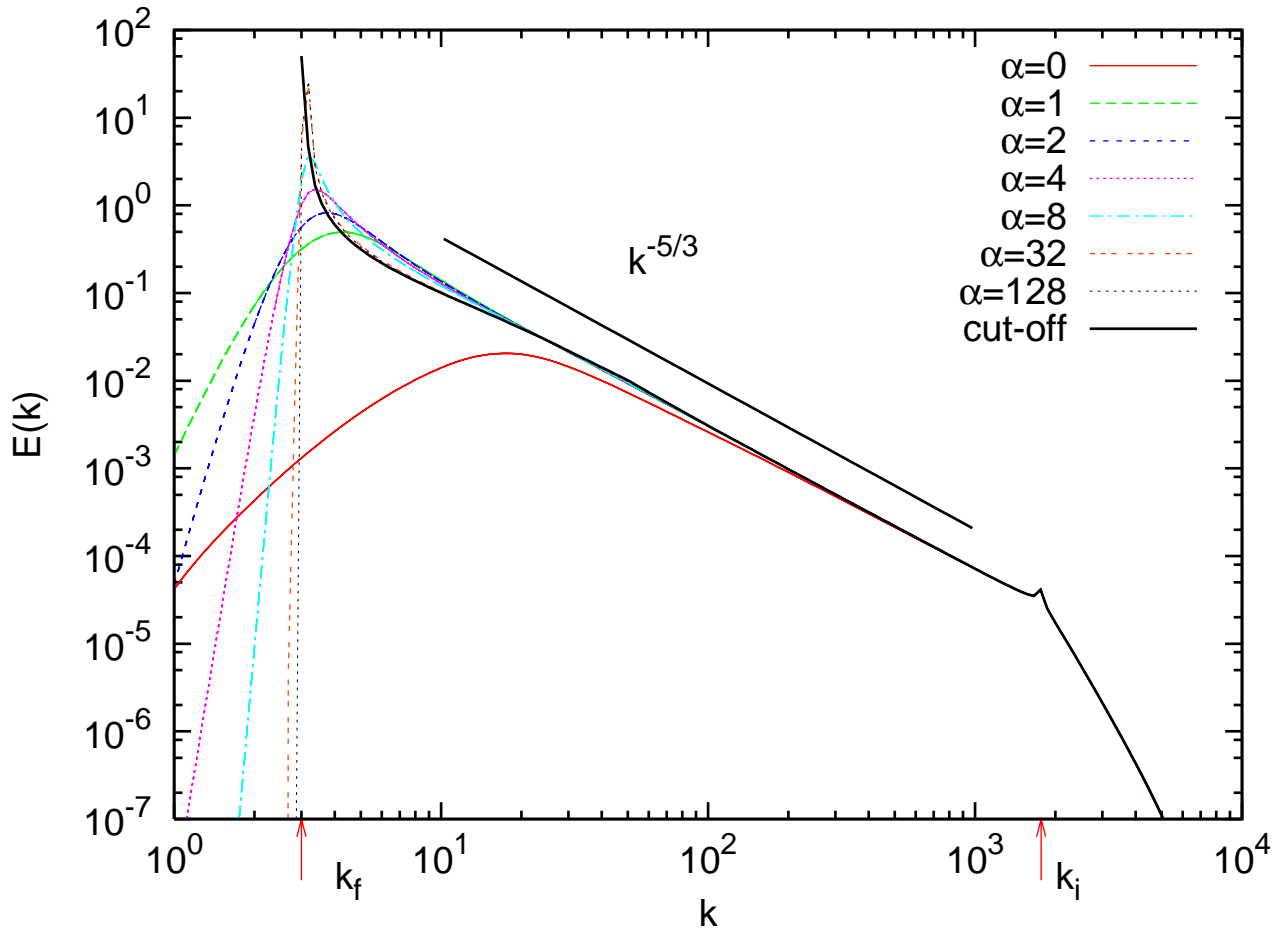}
\caption{Steady-state energy spectra with frictionality ranging from $0$ to
$128$ and a friction free case truncated at the low wavenumbers at
$k=k_f$. \label{fig1}}
\end{figure*}

\begin{figure*}
\setlength{\unitlength}{1\textwidth}
\includegraphics[width=1\unitlength]{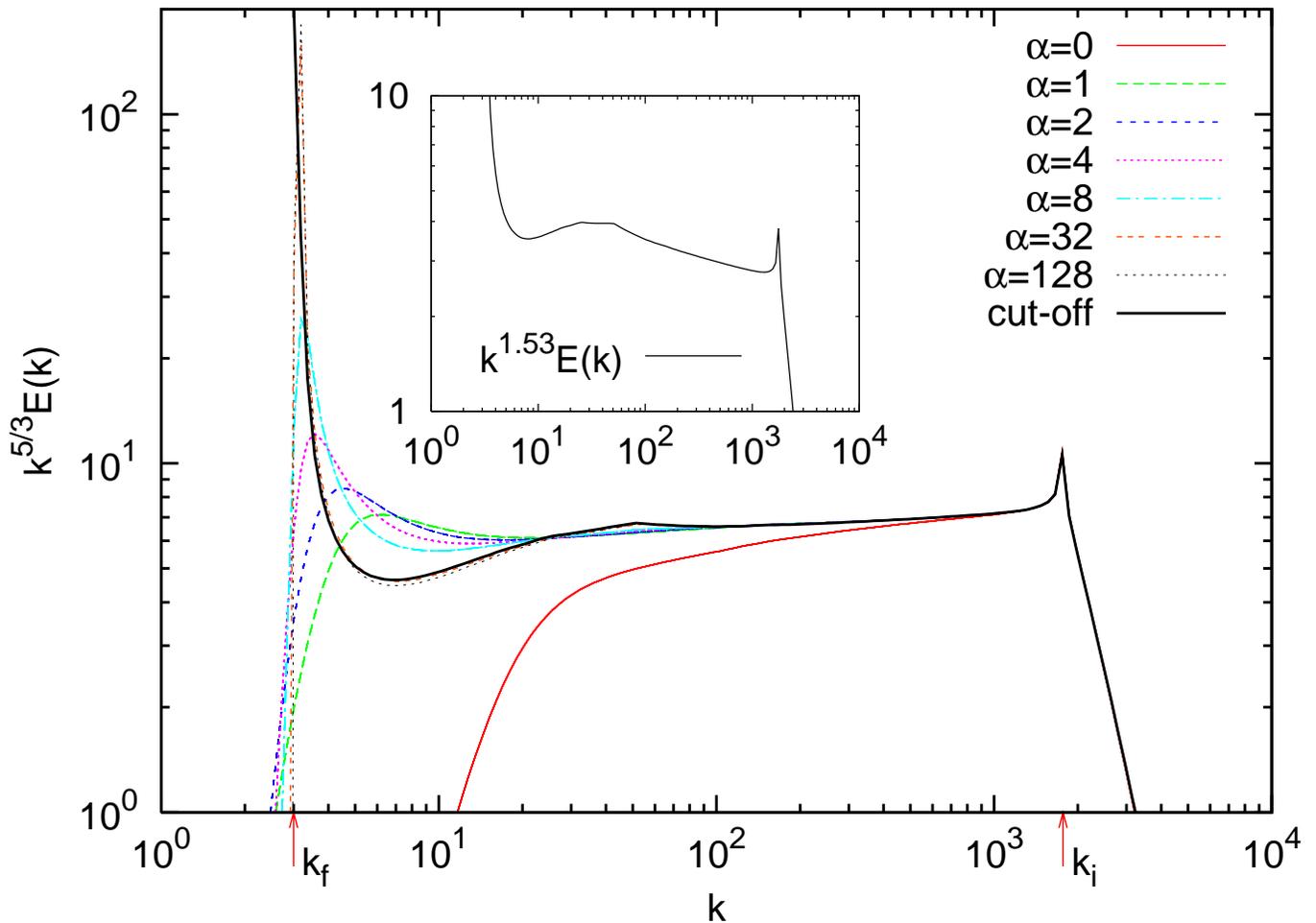}
%\begin{picture}(0,0)
%\put(0.25,0.36){\includegraphics[width=0.45\unitlength]{specsC32.eps}}
%\end{picture}
\caption{The same spectra as in Figure \ref{fig1} compensated by $k^{5/3}$ and, in the inset, compensated by $k^{1.53}$.  \label{fig2}}
\end{figure*}

For all runs but this last one, a steady state was obtained, for
which the energy spectrum is shown in Figure \ref{fig1}. For the
friction-free (truncated) case, a steady state will not be
obtained until the energy-pile up at the fundamental is balanced by
the viscous dissipation at this wavenumber, which is very small in
the present case as the Reynolds number is large.
The fundamental would then
accumulate an amount of energy which is enormous compared to the
cases with small frictionality. For a better comparison between the
different cases, in the following a spectrum is shown for which the
energy pile-up at the fundamental is comparable to the pile-up for
$\alpha=128$. In Figure \ref{fig2} we observe that for $\alpha>0$ a
bottleneck is present around $k_f\approx 3$. The size of this
bottleneck increases with $\alpha$. For values of $1\leq\alpha<8$, the bottleneck has the form of a bump. For $\alpha\geq 8$, the bottleneck changes its shape towards a peak and for the highest values of
$\alpha$ used in the present work, the bottleneck collapses with the
energy spectrum of the frictionless truncated case.  The $k^{-5/3}$ inertial
range is influenced by this bottleneck, not only on the bottleneck
itself, where the spectrum is steeper than $k^{-5/3}$, but also in
the region with wavenumbers larger than the bottleneck, and
extending for about one decade,  the spectrum deviates and is
slightly shallower than $k^{-5/3}$. The friction-free truncated case, and the high-$\alpha$ friction cases, show much similarity (at small $k$) with the inviscid truncated 3D case (at high $k$) studied in \cite{Bos2006-2}.
A depletion of the compensated energy spectrum is present in the vicinity of the peak (for $k>k_f$ in the 2D case, and for $k<k_{th}$ in the 3D case, where $k_{th}$ is the wavenumber marking the beginning of the equipartition range). The effect can, in both cases, be explained by nonlocal interactions, as was done in the three-dimensional case \cite{Bos2006-2}.

It is interesting to investigate whether this  depleted region in the compensated energy spectrum corresponds to the emergence of a new scaling regime. Dimensional analysis based on the assumption of a constant flux of energy and a sweeping timescale related to the advection of the small eddies by the velocity associated with the kinetic energy contained in the condensate, one can deduce a scaling proportional to $k^{-3/2}$. It is observed in the inset of Figure \ref{fig2} that a small wavenumber range shows a spectral exponent close to this value (1.53). This range is however too short to make conclusive statements.

It was shown by Smith and Yakhot \cite{Smith1993}, that when the condensate becomes large enough, the statistics of the velocity field become strongly non-Gaussian. One could then ask, if the present closure approach, based on an assumption of maximum randomness \cite{KraichnanDIA}, is still valid. To estimate the extend to which the closure gives results for the energy spectrum close to results of DNS, we compare our results with DNS results from literature \cite{Hossain1983}. In that study the coexistence of a large peak at the fundamental and a $k^{-5/3}$ range for larger wave-numbers is observed. The value of the energy accumulated at the fundamental is one order of magnitude above its inertial range value. In Figure \ref{fig2} the peak value is also one order of magnitude above its inertial range value. So at least qualitatively, the coexistence of a large peak and the $k^{-5/3}$ inertial range is observed in DNS as well as in closure. However this validation remains qualitative and the present results 
could only be fully supported by a detailed and quantitative comparison 
with DNS. Such a comparison is beyond the scope of the present paper.

When the bottleneck becomes large enough, a small
kink is observed in the spectrum around $k=50$. This comes from the
discretization. Nonlocal interactions are not resolved if the ratio of
the smallest wavenumber to the middle wavenumber of an interacting
triad is smaller than\cite{Pouquet} $a=2^{1/F}-1$. For elongated triads, the middle wavenumber and the longest wavenumber are approximately of the same
size. The largest wavenumber $k$ directly interacting with $k_f$ can then be shown to be $k\approx 50$ for $F=12$. This clearly illustrates the
nonlocal origin of the pseudo-friction range, created by the
bottleneck. In principle, the effect of the truncation can be removed
by refining the discretization. To allow nonlocal interactions over
the full $4$ decades of the simulation, the number of necessary gridpoints can
be estimated to be roughly $700$ per octave which would make the simulations
prohibitively expensive. Another solution would be to explicitly
compute the first-order contributions of the nonlocal interactions
\cite{Pouquet} by expanding the nonlinear transfer with respect to the
parameter $a$, which is small for very nonlocal interactions. This procedure has the disadvantage not to conserve
energy or to mix-up the different orders of the expansion. The
improvement expected to be minimal with respect to the presented
results, these solutions will not be attempted here.

\section{Conclusion}

The present results show that the bottleneck effect is not only
observable at the small scales, but it can also be created, be it
artificially, at the large scales. It shows that by increasing
$\alpha$, the energy spectrum tends to the condensed solution
predicted by Kraichnan \cite{Kraichnan67}. 
This constitutes a logical extension of the ideas of Frisch \etal  \cite{Frisch2008}
  to the inverse cascade in two-dimensional turbulence.
It also shows that
pile-up of energy at small wavenumbers influences the inertial range
scaling by nonlocal interactions.
 The present work thereby confirms the
ideas by Sukoriansky \etal \cite{Sukoriansky1999,Sukoriansky2005}, that large-scale
friction can influence all scales of two-dimensional turbulence,
since the pile-up of energy at the small scales might trigger the
creation of coherent structures. These coherent structures can
induce a $k^{-3}$ scaling\cite{Borue1994,Chertkov2007}, which
camouflates the inertial range. The use of hypofriction, intended to
enlarge the inertial range, can thus have the opposite effect.

A detailed DNS study would be necessary to confirm the effects observed in
the present study. Even at low or moderate resolution, trends could probably be observed and answer the question whether the predictions of the present closure, based on an expansion about Gaussianity, are valid for the flow here considered in which coherent large scale structures are present. In the authors' opinion, in this type of flow, as is the case for many types of turbulent fields, the simultaneous use of Direct Numerical Simulations at low and moderate Reynolds numbers and of closures at high Reynolds numbers will probably constitute the best strategy to gain understanding of the physics of turbulence dynamics. Such a study is however left for future work.

\section*{Acknowledgments}

We acknowledge interaction and discussion with Claude Cambon, Uriel Frisch, Fabien Godeferd, Walter Pauls, Samriddhi Sankar Ray, Jian-Zhou Zhu and two anonymous referees.

\end{document}